\documentclass[12pt]{iopart}
\usepackage[utf8]{inputenc}
\usepackage[T2A]{fontenc}
\usepackage[english]{babel}
\usepackage{bm}
\usepackage{graphicx}
\usepackage[sort]{cite}                 
\usepackage{booktabs,multirow,hhline}
\begin{document}

\title[]
{Gaussian polymer chains in a harmonic potential: The path integral approach}

\author{G V Paradezhenko$^1$, C Gascoigne$^2$ and N V Brilliantov$^{1}$ }
\address{$^1$ Center for Computational and Data-Intensive Science and Engineering,
Skolkovo Institute of Science and Technology, Moscow 121205, Russia}
\address{$^2$ Department of Mathematical Sciences, University of Bath, Claverton Down, Bath
BA2 7AY, United Kingdom}
\ead{g.paradezhenko@skoltech.ru}

\vspace{10pt}
\begin{indented}
\item[]14 August 2020
\end{indented}

\begin{abstract}
We study conformations of the Gaussian polymer chains in $d$-dimensional space in the presence of
an external field with the harmonic potential. We apply a path integral approach to derive
an explicit expression for the probability distribution function of the gyration radius.
We calculate this function using Monte Carlo simulations and
show that our numerical and theoretical results are in a good agreement
for different values of the external field.
\end{abstract}

%
\vspace{2pc}

\submitto{\jpa}
%
%
%


\section{Introduction}

The ideal (Gaussian) chain plays the same basic role in polymer theory as an ideal gas does
in the theory of gases \cite{FloryBook,khokhlov1994}.
Although the real polymers are very different from the Gaussian idealization, many
prominent properties of a polymer chain are adequately reflected within this idealized model. For instance, the Gaussian
chain may be used with an acceptable accuracy for estimating a chain configurational entropy; it is also a
convenient basic model in various thermodynamic perturbation theories (see, e.g., \cite{khokhlov1994,DE13}). Moreover, the
Gaussian chain model is used in many theories of conformational phase transitions, where the gyration radius of a polymer chain dramatically changes. This requires the dependence of the free energy of a chain as a function of its gyration radius. The total free energy is represented as a sum of free energies associated with different interactions
between the system constituents, as well as the free energy of a Gaussian chain with
a given gyration radius $R_g$
(see, e.g., \cite{GK92,BKK98,brilliantov2016,brilliantov2017,BudkovPolar2017,
budkov2017statistical,gordievskaya2018interplay,kolesnikov2017statistical,budkov2015flexible,budkov2017polymer,
budkov2016statistical}). Hence, it is important to have an accurate estimate for the free energy of an ideal chain with a
particular value of $R_g$. This quantity can be straightforwardly obtained from the probability distribution function
(PDF) of the gyration radius $P(R_g^2)$. Such a function
is a valuable tool for calculating important characteristics of the
chain; such as, powers of the gyration radius $\left< R_g^a \right>$ and other quantities needed
in the thermodynamic perturbation theory.

Generally, the statistical properties of the Gaussian polymer chains can be described within the Wang-Uhlenbeck
approach based on averaging of the polymer chain microscopic distribution function over all possible conformations
(see, e.g.,~\cite{DE13,Yam71}). An application of this method for calculating $P(R_g^2)$
was first proposed by
Fixman~\cite{Fix62}, who expressed the PDF in the form of a complex integral and obtained its asymptotic solutions for
small and large values of $R_g$. Later, the same results were obtained by a slightly different method
in~\cite{FH63,For65}. However, as was pointed out in~\cite{For65}, the asymptotic solution for large $R_g$ did not
perfectly agree with that of \cite{Fix62}.
A proper and accurate analytic solution of the Fixman integral was
reported in~\cite{FN70}. A similar method to the Fixman approach has been further developed in the recent studies
\cite{nech1,nech2}.

Given the numerous applications, it is important to know the PDF for the gyration radius of a Gaussian chain in an
external field. The nature of this field may be very different; for instance, it may be a field associated with the
interaction of a polymer with the solid matrix in a porous medium, or with other polymers in a polymer brush. Furthermore,
it may be an effective self-consistent field associated with the volume or other interactions between the chain
monomers. Using the PDF $P(R_g^2)$ in an external field, one can develop a more accurate thermodynamic perturbation theory
as compared to the theory based on the field-free PDF. The Gaussian chain in an external field has been investigated in \cite{KRC97} with the use of the stochastic differential equations for the bridge process. The energy distribution
function for chains in the harmonic external field was calculated;
however, the numerical or analytic
results for $P(R_g^2)$ are still lacking.

In the present study, we analyze the conformation of $d$-dimensional Gaussian chains in a harmonic potential by means
of the path integral  approach that is a rather popular method in polymer physics (see, e.g.,~\cite{DE13,Vil00,Kle04}).
An application of this method for calculating the PDF $P(R_g^2)$ for a field-free case was proposed in \cite{BK16},
where the Fixman's results were reproduced. Here we further develop this approach and obtain new results for $P(R_g^2)$
in the external field. We wish to stress that finding $P(R_g^2)$ in the external potential 
by the Fixman method is not straightforward, or even possible.
This is in contrast to the path integral approach proving its flexibility and universality. 
Using $P(R_g^2)$, we obtain an analytic expression for the free energy of
a Gaussian chain in an external potential as a function of the gyration radius and potential strength. 
Such quantities are useful in the thermodynamic perturbation theories. 
To check the predictions of the theory, we perform Monte Carlo
simulations and demonstrate a very good agreement between theoretical and numerical findings.

This paper is organized as follows. In Sec.~\ref{sect:prob-Rg}, we express the PDF $P(R_g^2)$ in the harmonic
potential in terms of the path integral.
In Sec.~\ref{sect:charfun}, the characteristic function of $P(R_g^2)$ is
calculated with the use of the saddle-point method. In Sec.~\ref{sect:PDF},
we derive analytic expressions for the PDF
in the limits of small and large expansion factor.
Sec.~\ref{sect:numres} is devoted to Monte Carlo simulations; here we
compare our theory with the numerical results. Finally, Sec.~\ref{sect:conclusion} summarizes our findings.

\section{The probability distribution function of the gyration radius}
\label{sect:prob-Rg}

The partition function of Gaussian polymer chains in the presence of an external field may be written in terms of
a path integral as (see, e.g.,~\cite{Vil00,DE13})
\begin{equation}\label{Z-def}
  Z = \int \mathcal{D} \mathbf{r}\, e^{-S_0[\mathbf{r}]},
\end{equation}
where the functional $S_0[\mathbf{r}]$ is given by
\begin{equation}\label{S0-def}
  S_0[\mathbf{r}]
  = \frac{d}{2 b^2} \int_0^N d \tau \, \dot{\mathbf{r}}^2(\tau)
  + \frac{1}{k_{\mathrm{B}}T} \int_0^N d \tau \, V(\mathbf{r}(\tau)).
\end{equation}
Here, $N$ is the number of monomers in the chain, $d$ is the dimension of space, $b$ is the Kuhn length of the segment,
$k_\mathrm{B}$ is the Boltzmann constant, $T$ is the temperature  and $V(\mathbf{r})$ is the potential of the external
field. We assume the potential is harmonic:
\begin{equation}\label{V-field-potential}
  V(\mathbf{r}) = \frac{\Omega^2}{2} (\mathbf{r}  - \mathbf{r}_0)^2,
\end{equation}
where $\Omega$ characterizes the steepness of the parabolic well and $\mathbf{r}_0$ is the fixed point.
Substituting Eq. (\ref{V-field-potential}) in
(\ref{S0-def}) and introducing the new variable $s = \tau/N$, we obtain
\begin{equation}\label{S0-harmonic}
  S_0[\mathbf{r}]
  = \frac{c}{2} \int_0^1 d s \,
  \left[ \dot{\mathbf{r}}^2(s) + \lambda^2 \left(\mathbf{r}(s) - \mathbf{r}_0 \right)^2 \right],
\end{equation}
where $c = d/(N b^2)$ and
\begin{equation}\label{lambda-def}
  \lambda^2  =  \frac{\Omega^2 N^2 b^2}{d k_{\mathrm{B}}T}.
\end{equation}

To study the distribution of the gyration radius $R_g$, we consider the conditional partition function (\ref{Z-def}) at
fixed $R_g$:
\begin{equation}\label{Z-Rg}
  Z(R_g^2) =
  \int \mathcal{D} \mathbf{r} \, e^{-S_0[\mathbf{r}]} \,
  \delta \left( R_g^2 - \frac{1}{2} \int_0^1 \int_0^1  d s_1 d s_2  \,
  [\mathbf{r}(s_1) - \mathbf{r}(s_2)]^2 \right).
\end{equation}
Hence the PDF of $R_g^2$ reads,
\begin{equation}\label{PDF-def}
  P(R_g^2) \equiv Z(R_g^2)/Z.
\end{equation}
From Eqs. (\ref{Z-Rg}) and (\ref{PDF-def}), it is clear the PDF is normalized,
$\int d R_g^2 \, P(R_g^2) = 1$. Now, substituting the representation of the delta function
\begin{equation*}
  \delta(x) =  \frac{1}{2\pi}\int d \xi \, e^{-i x \xi},
\end{equation*}
into  Eq. (\ref{Z-Rg}), we write the PDF in Eq. (\ref{PDF-def}) as
\begin{equation}\label{PDF-fourier}
  P(R_g^2) = \frac{1}{2\pi} \int d \xi \, K(\xi)\, e^{-i \xi R_g^2},
\end{equation}
where
\begin{equation}\label{K-def}
  K(\xi)
  = \frac{1}{Z}
 \int \mathcal{D}\mathbf{r} \,
  \exp \left( -S_0[\mathbf{r}]
  + \frac{i \xi}{2} \int_0^1 \int_0^1  d s_1 d s_2 \,
  [\mathbf{r}(s_1) - \mathbf{r}(s_2)]^2 \right)
\end{equation}
is the standard characteristic function (see, e.g., \cite{Pap91}). Thus, the calculation of $P(R_g^2)$ involves two
steps. First, we compute the characteristic function $K(\xi)$, given by the path integral (\ref{K-def}).
Second, the
inverse Fourier transform (\ref{PDF-fourier}) yields the PDF $P(R_g^2)$.

\section{Calculation of the characteristic function}
\label{sect:charfun}

To recast $K(\xi)$, given by Eq.  (\ref{K-def}), into a more convenient form, we expand the second term in the
exponential expression of (\ref{K-def}):
\begin{eqnarray*}
  \frac{i \xi}{2}\int_0^1 \int_0^1  d s_1 d s_2 \,
  [\mathbf{r}(s_1) - \mathbf{r}(s_2)]^2 \\
  =
  \frac{i \xi}{2}
  \left( \int_0^1  d s_1 \, \mathbf{r}^2(s_1) + \int_0^1  d s_2 \, \mathbf{r}^2(s_2)
  - 2\int_0^1 d s_1  \, \mathbf{r}(s_1) \int_0^1 d s_2 \, \mathbf{r}(s_2) \right).
\end{eqnarray*}
Making the substitution $s_1, s_2 \mapsto s$ (recall that $s_1$ and $s_2$ are dummy variables), we obtain
\begin{equation}\label{s1s2-s}
  \fl \frac{i \xi}{2}\int_0^1 \int_0^1  d s_1 d s_2 \,
  [\mathbf{r}(s_1) - \mathbf{r}(s_2)]^2
  = i \xi
  \left[ \int_0^1  d s \, \mathbf{r}^2(s)
  - \left( \int_0^1 d s  \, \mathbf{r}(s) \right)^2 \right],
\end{equation}
which yields,
\begin{equation}\label{K-1}
  K(\xi)
  = \frac{1}{Z}
  \int \mathcal{D}\mathbf{r} \,
  \exp \left[ - S_0[\mathbf{r}] + i\xi \int_0^1 d s \, \mathbf{r}^2(s)
  - i\xi \left( \int_0^1 d s  \, \mathbf{r}(s)  \right)^2 \right].
\end{equation}
Applying the well-known integral relation
\begin{equation*}
  \exp\left(-\frac{b^2}{4a}\right) = \sqrt{\frac a\pi} \int d x \, \exp\left( -a x^2 + i b x \right)
\end{equation*}
with $a = 1/(4i\xi)$ and $b = \int_0^1 d s  \, \mathbf{r}(s) $ to (\ref{K-1}), we obtain
\begin{equation}\label{K-2}
  K(\xi) = \frac{1}{Z} \left( \frac{1}{4\pi i\xi} \right)^{\frac d2}
  \int d \mathbf{x}\, e^{\frac{i  \mathbf{x}^2}{4\xi}}
  \int \mathcal{D}\mathbf{r} \, e^{-S[\mathbf{r}]},
\end{equation}
with
\begin{equation*}
  S[\mathbf{r}] = S_0[\mathbf{r}] - i\xi \int_0^1 d s \, \mathbf{r}^2(s)
  - i \mathbf{x} \int_0^1 d s \, \mathbf{r}(s).
\end{equation*}
Using Eq. (\ref{S0-harmonic}) and introducing the following notation
\begin{equation}\label{omega-def}
  \omega^2 = \frac{2i \xi}{c} - \lambda^2,
\end{equation}
we recast $S[\mathbf{r}]$ into the form
\begin{equation}\label{S-def}
  S[\mathbf{r}] = \frac{c}{2} \int_0^1 d s \left[ \dot{\mathbf{r}}^2(s) - \omega^2 \mathbf{r}^2(s) \right]
  - i \mathbf{y} \int_0^1 d s \, \mathbf{r}(s) + \frac{c\lambda^2}{2} \mathbf{r}_0^2,
\end{equation}
where $\mathbf{y} = \mathbf{x} - i c\lambda^2 \mathbf{r}_0$.

Further development of the characteristic function $K(\xi)$ requires the calculation of the Gaussian path integrals in
Eqs. (\ref{Z-def}) and (\ref{K-2}).
These path integrals can be computed by means of the saddle-point method;
in the case of a Gaussian integral, it gives an exact result (see, e.g.,~\cite{Zin05}).
We specify the boundary conditions as
follows. We first assume that one of the ends is fixed at the origin, $\mathbf{r}(0) = 0$,
and then assume $\mathbf{r}(1) = \mathbf{R}$. In order to take into account all possible conformations of the chains, we integrate over $\mathbf{R}$. Hence, expression (\ref{K-2}) takes the form
\begin{equation}\label{K-3}
  K(\xi) = \frac{1}{Z} \left( \frac{1}{4\pi i\xi} \right)^{\frac d2}
  \int d \mathbf{x}\, e^{\frac{i  \mathbf{x}^2}{4\xi}}
  \int d \mathbf{R} \,
  \int_{\mathbf{r}(0) = 0}^{\mathbf{r}(1) = \mathbf{R}} \mathcal{D}\mathbf{r} \, e^{-S[\mathbf{r}]},
\end{equation}
and for normalization (\ref{Z-def}) we have
\begin{equation}\label{Z-1}
  Z = \int d \mathbf{R} \,
  \int_{\mathbf{r}(0) = 0}^{\mathbf{r}(1) = \mathbf{R}} \mathcal{D}\mathbf{r} \, e^{-S_0[\mathbf{r}]}.
\end{equation}

To compute $Z$ by the saddle point method, we write $\mathbf{r}(s) = \mathbf{u}(s) + \bm{\rho}(s)$, where the extremal path $\mathbf{u}(s)$ satisfies the Euler-Lagrange equation for functional (\ref{S0-harmonic}):
\begin{equation}\label{sp-equation-S0}
  \frac{\delta S_0[\mathbf{u}]}{\delta \mathbf{u}(s)}
  = c \left[ \ddot{\mathbf{u}}(s) - \lambda^2 \mathbf{u}(s) + \lambda^2 \mathbf{r}_0 \right] = 0,
\end{equation}
with the boundary conditions $\mathbf{u}(0) = 0$ and $ \mathbf{u}(1) = \mathbf{R}$. The solution to differential
equation (\ref{sp-equation-S0}) reads
\begin{equation}\label{sp-u}
  \mathbf{u}(s)
  = \mathbf{r}_0 \left( 1 - e^{-\lambda s} - \frac{1 - e^{-\lambda}}{\sinh \lambda} \sinh(\lambda s) \right)
  + \mathbf{R}\, \frac{ \sinh(\lambda s) }{\sinh \lambda}.
\end{equation}
The function $\bm{\rho}(s)$ describes random fluctuations near the saddle-point and satisfies the uniform boundary
conditions: $\bm{\rho}(0) = \bm{\rho}(1) = 0$. Since
\begin{equation*}
  S_0[\mathbf{r}] = S_0[\mathbf{u} + \bm{\rho}]
  =  S_0[\mathbf{u}]
  + \frac c2 \int_0^1 d s\, \left[ \dot{\bm{\rho}}^2(s) + \lambda^2 \bm{\rho}^2(s) \right],
\end{equation*}
the expression (\ref{Z-1}) takes the form
\begin{equation}\label{Z-2}
  Z
  =
  \int_{\bm{\rho}(0) = 0}^{\bm{\rho}(1) = 0} \mathcal{D}\bm{\rho}\,
  \exp\left[ -\frac c2 \int_0^1 d s\, \left[ \dot{\bm{\rho}}^2(s) + \lambda^2 \bm{\rho}^2(s) \right] \right] \,
  \int d\mathbf{R}\, e^{-S_0[ \mathbf{u}]}.
\end{equation}
Integrating by parts and taking into account Eq. (\ref{sp-equation-S0}),
we evaluate $S_0[\mathbf{u}]$ as
\begin{equation*}
  S_0[\mathbf{u}]
  = \frac c2 \left[ \mathbf{u}(1) \dot{\mathbf{u}}(1) -
  \lambda^2 \mathbf{r}_0 \int_0^1 d s \,\mathbf{u}(s) + \lambda^2 \mathbf{r}_0^2 \right].
\end{equation*}
Using (\ref{sp-u}), we arrive at
\begin{equation*}
  S_0[\mathbf{u}]
  = \frac c2 \left[ \frac{\lambda}{\tanh \lambda} \, \mathbf{R}^2
  - 2\lambda \tanh \left(\frac{\lambda}{2} \right) \mathbf{R}\mathbf{r}_0
  + 2\lambda \tanh \left(\frac{\lambda}{2} \right) \mathbf{r}_0^2 \right].
\end{equation*}
Substituting the latter result into (\ref{Z-2}) and calculating the Gaussian integral over $\mathbf{R}$, we find
\begin{equation}\label{Z-fin}
  \fl Z =
  \left( \frac{2\pi \tanh \lambda}{c \lambda} \right)^{\frac d2}
  e^{-\mathbf{r}_0^2  \frac{c\lambda}{2} \tanh \lambda}
  \int_{\bm{\rho}(0) = 0}^{\bm{\rho}(1) = 0} \mathcal{D}\bm{\rho}\,
  \exp\left[ -\frac c2 \int_0^1 d s\, \left[ \dot{\bm{\rho}}^2(s) + \lambda^2 \bm{\rho}^2(s) \right] \right].
\end{equation}

Similarly, we calculate the Gaussian path integral in Eq. (\ref{K-2}) for the characteristic function. Assuming
$\mathbf{r}(s) = \mathbf{v}(s) + \bm{\rho}(s)$, we write the Euler-Lagrange equation for the functional (\ref{S-def}):
\begin{equation}\label{sp-equation-S}
  \frac{\delta S[\mathbf{v}]}{\delta \mathbf{v}(s)}
  = c \left[ \ddot{\mathbf{v}}(s) + \omega^2 \mathbf{v}(s) \right] + i \mathbf{y} = 0.
\end{equation}
The solution to differential equation (\ref{sp-equation-S}) with
the boundary conditions $\mathbf{v}(0) = 0$ and $\mathbf{v}(1) = \mathbf{R}$ reads
\begin{equation}\label{sp-v}
 \mathbf{v}(s)
  = -\frac{i \mathbf{y}}{\omega^2 c} \left(1 - \cos(\omega s)
  - \frac{1 - \cos \omega}{\sin\omega} \sin(\omega s) \right)
  + \frac{\mathbf{R}}{\sin \omega} \sin(\omega s).
\end{equation}
Integrating by parts in (\ref{S-def}) and taking into account (\ref{sp-equation-S}),
we recast $S[\mathbf{v}]$ into the form
\begin{equation*}
  S[\mathbf{v}]
  = \frac c2 \, \mathbf{v}(1) \dot{\mathbf{v}}(1)
  - \frac{i \mathbf{y}}{2} \int_0^1 d s \, \mathbf{v}(s) + \frac{c\lambda^2}{2} \mathbf{r}_0^2.
\end{equation*}
Substituting now the solution for $\mathbf{v}(s)$ from Eq. (\ref{sp-v}), we arrive at
\begin{equation}\label{S-v-sp}
  S[\mathbf{v}]  = \frac{c\mathbf{R}^2 \omega}{2\tan \omega}
  - \frac{i \mathbf{y} \mathbf{R}}{\omega}\tan\left(\frac \omega 2\right)
  + \frac{\mathbf{y}^2}{c \omega^3} \tan\left(\frac \omega 2\right)
  - \frac{\mathbf{y}^2}{2 c \omega^2} + \frac{c\lambda^2}{2} \mathbf{r}_0^2.
\end{equation}
Finally, taking into account that
\begin{equation*}
  S[\mathbf{r}] = S[\mathbf{v} + \bm{\rho}]  =  S[\mathbf{v}]
  + \frac c2 \int_0^1 d s\, \left[ \dot{\bm{\rho}}^2(s) - \omega^2 \bm{\rho}^2(s) \right]
\end{equation*}
and substituting (\ref{Z-fin}) into (\ref{K-3}), we obtain
\begin{equation}\label{K-4}
  \fl K(\xi)
  =
  \left( \frac{c \lambda}{2\pi \tanh \lambda} \right)^{\frac d2}
  e^{\mathbf{r}_0^2  \frac{c\lambda}{2} \tanh \lambda}
  \left( \frac{1}{4\pi i\xi} \right)^{\frac d2}
  G(\omega,\lambda) \int d \mathbf{x}\, e^{\frac{i  \mathbf{x}^2}{4\xi}}
  \int d \mathbf{R} \, e^{-S[\mathbf{v}]},
\end{equation}
where
\begin{equation}\label{G-def}
  G(\omega,\lambda)
  = \frac{\int_{\bm{\rho}(0) = 0}^{\bm{\rho}(1) = 0} \mathcal{D}\bm{\rho}
  \exp\left[ -(c/2)\int_0^1 d s \,\left( \dot{\bm{\rho}}^2 - \omega^2 \bm{\rho}^2 \right) \right]}
  {\int_{\bm{\rho}(0) = 0}^{\bm{\rho}(1) = 0} \mathcal{D}\bm{\rho}\,
  \exp\left[ -(c/2)\int_0^1 d s \,\left( \dot{\bm{\rho}}^2 + \lambda^2 \bm{\rho}^2 \right) \right]}.
\end{equation}

To compute $G(\omega,\lambda)$, we follow \cite{BK16} and expand the functions $\bm{\rho}(s)$ in the Fourier series
$\bm{\rho}(s) = \sum_{n=1}^{\infty} \bm{\rho}_n \sin (\pi n s)$.
Rewriting the integrals in (\ref{G-def}) in terms of the Fourier components $\bm{\rho}_n$, yields
\begin{equation*}
  G(\omega,\lambda)
  = \frac{\int \prod_{n=1}^{\infty} d \bm{\rho}_n \exp\left[ -(c/2) (\pi^2n^2 - \omega^2) \bm{\rho}_n^2 \right]}
  {\int \prod_{n=1}^{\infty} d \bm{\rho}_n \exp\left[ -(c/2) (\pi^2n^2 + \lambda^2) \bm{\rho}_n^2 \right]}.
\end{equation*}
Calculating both Gaussian integrals in the numerator and denominator,
we obtain
\begin{equation*}
  G(\omega,\lambda)
  = \prod_{n=1}^{\infty} \left( \frac{ \pi^2n^2 + \lambda^2 }{ \pi^2n^2 - \omega^2 } \right)^{\frac d2}
  = \prod_{n=1}^{\infty} \left(
  \frac{1-\frac{(i \lambda)^2}{\pi^2n^2}}{1 -\frac{\omega^2}{\pi^2n^2}}
  \right)^{\frac d2}.
\end{equation*}
With the infinite product representation (see, e.g., \cite{Mel12})
\begin{equation*}
  \sin x = x \prod_{n=1}^{\infty}\left( 1 - \frac{x^2}{\pi^2 n^2} \right),
\end{equation*}
we come to the result
\begin{equation}\label{G-fin}
  G(\omega,\lambda)
  = \left( \frac{\sinh \lambda}{\lambda} \right)^{\frac d2}
  \left( \frac{\omega}{\sin \omega} \right)^{\frac d2}.
\end{equation}

Next, we proceed to the final step in the calculations of the characteristic function.
Substituting (\ref{G-fin}) into (\ref{K-4}), we obtain
\begin{equation}\label{K-5}
  \fl K(\xi)
  =
  \left( \frac{c \cosh \lambda}{2\pi} \right)^{\frac d2}
  e^{\mathbf{r}_0^2  \frac{c\lambda}{2} \tanh \lambda}
  \left( \frac{\omega}{\sin \omega} \right)^{\frac d2}
  \left( \frac{1}{4\pi i\xi} \right)^{\frac d2}
  \int d \mathbf{x}\, e^{\frac{i  \mathbf{x}^2}{4\xi}}
  \int d \mathbf{R} \, e^{-S[\mathbf{v}]}.
\end{equation}
To calculate the Gaussian integral on the right-hand side of Eq. (\ref{K-5}),
we make the substitution $\mathbf{x} = \mathbf{y} + i c \lambda^2 \mathbf{r}_0$:
\begin{equation*}
  I = \left( \frac{1}{4\pi i\xi} \right)^{\frac d2}
  \int d \mathbf{y}\, \exp\left[ \frac{i  (\mathbf{y} + i c \lambda^2 \mathbf{r}_0)^2}{4\xi} \right]
  \int d \mathbf{R} \, e^{-S[\mathbf{v}]}.
\end{equation*}
Using Eqs. (\ref{lambda-def}) and (\ref{S-v-sp}), we evaluate this Gaussian integral as
\begin{equation}\label{I-fin}
  I = \left( \frac{2 \pi}{c} \right)^{\frac d2}
  \left( \frac{1}{1 + \lambda^2 f(\omega)} \right)^{\frac d2}
  \exp\left[ - \frac{c\lambda^2 \mathbf{r}_0^2}{2(1 + \lambda^2 f(\omega))} \right],
\end{equation}
where
\begin{equation}\label{f-omega-def}
  f(\omega) = \frac{\tan \omega - \omega}{\omega^2 \tan \omega}.
\end{equation}
Substituting expression (\ref{I-fin})
into Eq. (\ref{K-5}), we arrive at the exact expression for the characteristic function
\begin{equation}\label{K-fin}
  \fl K(\xi) =
  \exp\left[\frac{c\lambda \mathbf{r}_0^2}{2}
  \left( \tanh \lambda - \frac{\lambda}{1 + \lambda^2 f(\omega)} \right) \right]
  \left( \frac{\cosh \lambda}{1 +\lambda^2 f(\omega)} \right)^{\frac d2}
  \left( \frac{\omega}{\sin \omega} \right)^{\frac d2},
\end{equation}
where $\omega$ depends on $\xi$ according to (\ref{omega-def}).
For practical applications, it is worth considering when $\mathbf{r}_0=0$; what we
have done in the following.

\section{Calculation of the probability distribution function}
\label{sect:PDF}

Now we are in a position to compute the PDF of the gyration radius.
Using  Eq. (\ref{f-omega-def}), one can recast the characteristic function (\ref{K-fin}) into the
form
\begin{equation}\label{K-zero-r}
  K(\xi) = \left( \cosh \lambda \right)^{d/2}
  \left( \frac{\omega^3}{\omega^2 \sin \omega + \lambda^2 (\sin \omega - \omega \cos \omega)} \right)^{\frac d2}.
\end{equation}
Therefore, substituting (\ref{K-zero-r}) to (\ref{PDF-fourier}) and using
$i\xi = c(\omega^2 + \lambda^2)/2$, we come to
\begin{equation}\label{P-W-xi}
  P(R_g^2)
  = \frac{e^{-d \alpha^2 \lambda^2/12} \left( \cosh \lambda \right)^{d/2}}{2\pi}  \,
   \int d \xi \, e^{-W(\xi)},
\end{equation}
where $\alpha = R_g/\langle R_g^2 \rangle_0^{1/2}$ is the chain expansion factor,
$\langle R_g^2 \rangle_0^{} = Nb^2/6$ is the mean squared gyration radius of an ideal chain in the lack of an external field, and
\begin{equation}\label{W-2}
  W(\xi) = \frac{d \alpha^2 \omega^2}{12} - \frac{3d}{2}\ln \omega
  + \frac{d}{2} \ln \left(\omega^2 \sin \omega + \lambda^2 (\sin \omega - \omega \cos \omega) \right).
\end{equation}
For the field-free case, $\lambda =0$, the integral in Eq. (\ref{P-W-xi}) for $d=3$
may be expressed in terms of the infinite series \cite{FN70}:
\begin{eqnarray}\label{PDF-case-1}
  P(R_g^2) & = & \frac{1}{2\sqrt{2}\pi \alpha^7 \langle R_g^2 \rangle_0}
  \sum_{n=0}^{\infty} \frac{(2n+1)!}{(2^n n!)^2} (4n+3)^{\frac 72} \, e^{-t_n} \nonumber \\
  & \times & \left[ \left( 1 - \frac{5}{8 t_n}\right) K_{\frac 14}(t_n)
  + \left( 1 - \frac{3}{8 t_n}\right) K_{\frac 34}(t_n) \right],
\end{eqnarray}
where $t_n = (4n+3)^2/(8 \alpha^2)$ and $K_s(t)$ is the modified Bessel function
of the second kind. Although exact,
the above relation is not very practical and it would be worth to consider
the limiting cases of small $\alpha \ll 1$
and large $\alpha \gg 1$ expansion factor.

For the non-zero field case, we evaluate integral (\ref{P-W-xi}) approximately
for small $\alpha \ll 1$ and large $\alpha \gg 1$ expansion factor
by means of the saddle-point method (see, e.g., \cite{Zin05}). The idea is
to find the saddle-point of $W(\xi)$, where $W'(\xi_0) = 0$ and $W(\xi)$ is expanded in a Taylor
series,
\begin{equation}\label{W-approx}
  W(\xi) \approx W(\xi_0) + \frac{1}{2} W''(\xi_0)(\xi - \xi_0)^2.
\end{equation}
This approximation reduces (\ref{P-W-xi}) into a Gaussian integral.
Calculating the first derivative of $W(\xi)$ from Eq. (\ref{W-2}),
we can find the saddle-point equation explicitly,
\begin{equation}\label{saddle-point-eq}
\frac{ \alpha^2}{3} = \frac{3}{\omega^2}
  -  \frac{(2+\lambda^2) + \omega \cot \omega}
  {\omega^2  + \lambda^2 (1- \omega \cot \omega )}.
\end{equation}

We start from the limit $\alpha \ll 1$. As it follows from Eq. (\ref{saddle-point-eq}),
the limit $\alpha \to 0$ implies
$|\omega| \to \infty$. We seek the solution in the form $\omega(\xi_0) = i x$,
where $x \gg 1$. For any limited value of
$\lambda$, when $\lambda^2/x \to 0$ for $x \to \infty $ we obtain
\begin{equation}\label{saddle-point-eq-1}
  \frac{\alpha^2}{3}
  \simeq -\frac{3}{x^2} +\frac{1}{x}.
\end{equation}
The same equation (\ref{saddle-point-eq-1}) may be obtained for $\lambda \to \infty$,
which yields the solution of the
saddle-point equation that does not depend on $\lambda$, $\omega(\xi_0) \simeq 3i/\alpha^2$.
Evaluating (\ref{W-2}) at $\omega = 3i/\alpha^2$, we obtain
\begin{equation}\label{W-w0-final}
  W(\xi_0)
  \simeq
  \frac{3 d }{4\alpha^2} + d \ln \alpha
  - \frac{d}{2} \ln \left( \frac{18}{3 + \alpha^2 \lambda^2} \right).
\end{equation}
Evaluating the second derivative of $W(\xi)$ at the same point yields
\begin{equation}\label{d2W-w0-final}
  W''(\xi_0)
  \simeq
  \frac{d\alpha^6}{54 c^2} = \frac{N^2b^4 \alpha^6}{54d}.
\end{equation}
Using (\ref{W-w0-final}) and (\ref{d2W-w0-final}), we can write Eq. (\ref{W-approx}) as
\begin{equation}\label{W-approx-final}
  W(\xi) \approx
  \frac{3 d }{4\alpha^2} + d \ln \alpha
  - \frac{d}{2} \ln \left( \frac{18}{3 + \alpha^2 \lambda^2} \right)
  + \frac{N^2b^4 \alpha^6}{108d} \left( \xi - \xi_0\right)^2.
\end{equation}
Finally, we calculate the integral in Eq. (\ref{P-W-xi}), using the approximation (\ref{W-approx-final}):
\begin{eqnarray*}
  \fl P(R_g^2) & = & \frac{\left(\cosh \lambda \right)^{d/2}}{2\pi}
  \exp\left[ - \frac{d\alpha^2 \lambda^2 }{12}  -\frac{3 d }{4\alpha^2}  - d \ln \alpha
  + \frac{d}{2} \ln \left( \frac{18}{3 + \alpha^2 \lambda^2} \right) \right] \\
  \fl  & \times &
  \int d \xi \, \exp\left[ -\frac{N^2b^4 \alpha^6}{108d} \left( \xi - \xi_0\right)^2 \right].
\end{eqnarray*}
Calculating the above Gaussian integral, we arrive at the  PDF of the gyration radius for
a small expansion, $\alpha \ll 1$,
\begin{equation}\label{PDF-case-2}
  P(R_g^2) =
  \sqrt{\frac{3d}{4\pi}} \frac{\alpha^{-d-3}}
  {\langle R_g^2 \rangle_0}
  \left( \frac{18 \cosh \lambda }{3 + \alpha^2\lambda^2} \right)^{\frac d2}
  \exp \left[ -\frac{3d}{4\alpha^2} - \frac{d\alpha^2\lambda^2}{12} \right].
\end{equation}

Next we consider the case of a large expansion, $\alpha \gg  1$.
Physically, the large expansion $\alpha \gg 1$ is not relevant in the strong constraining field.
The configurations with large $\alpha$ in a strong field are very improbable;
this case requires a special analysis and will not be addressed here.
Therefore, one can consider the saddle-point equation (\ref{saddle-point-eq}) only for $\lambda\ll 1$
and recast it into the simplified form
\begin{equation}\label{saddle-point-eq-3}
  \frac{\alpha^2}{3} \simeq \frac{1}{\omega^2} \left( 1 - \frac{\omega}{\tan \omega} \right).
\end{equation}
The analysis of Eq. (\ref{saddle-point-eq-3}) suggests that to fulfill
the condition $\alpha \gg 1 $, we need to seek
the solution in the form $\omega(\xi_0) = \pi - x$ with $x \ll 1$. Substituting the latter
into Eq. (\ref{saddle-point-eq-3}) and keeping only the relevant terms, we obtain
$x = 3/(\pi\alpha^2)$. Thus, the solution of the saddle-point equation reads,
$\omega(\xi_0) = \pi - 3/(\pi\alpha^2)$.
Performing the same steps as for the previous case of $\alpha \ll 1$,
we arrive at the distribution function for $\alpha \gg  1$:
\begin{equation}\label{PDF-case-3}
  P(R_g^2) =
  \frac{\sqrt{d}}{2\pi^{1/2 - d}} \frac{\alpha^{d-2}}
  {\langle R_g^2 \rangle_0}
  \left( \frac{e \cosh \lambda }{3 + \alpha^2\lambda^2} \right)^{\frac d2}
  \exp \left[ -\frac{\pi^2 d \alpha^2}{12} - \frac{d\alpha^2\lambda^2}{12} \right].
\end{equation}

The above two expressions for the PDF obtained for a small $\alpha \ll 1$ (\ref{PDF-case-2})
and large $\alpha \gg 1$ (\ref{PDF-case-3}) expansion factor may be summarized as
\begin{equation}\label{PDF-fin}
  \fl P(R_g^2) = \cases{
  \sqrt{\frac{3d}{4\pi}} \frac{\alpha^{-d-3}}
  {\langle R_g^2 \rangle_0}
  \left( \frac{18 \cosh \lambda }{3 + \alpha^2\lambda^2} \right)^{\frac d2}
  \exp \left[ -\frac{3d}{4\alpha^2} - \frac{d\alpha^2\lambda^2}{12} \right],
  \quad \alpha \ll 1, \\
  \frac{\sqrt{d}}{2\pi^{1/2 - d}} \frac{\alpha^{d-2}}
  {\langle R_g^2 \rangle_0}
  \left( \frac{e \cosh \lambda }{3 + \alpha^2\lambda^2} \right)^{\frac d2}
  \exp \left[ -\frac{\pi^2 d \alpha^2}{12} - \frac{d\alpha^2\lambda^2}{12} \right],
  \quad \alpha \gg 1,
  }
\end{equation}
where the second formula for $\alpha \gg 1$ is valid only for a small field, $\lambda \ll 1$.

Note that sometimes a distribution function $P(R_g)$ is considered, instead of $P(R_g^2)$, which are simply related as
$P(R_g)= 2 R_g P(R_g^2) $ (see, e.g., \cite{Yam71}).

Taking the logarithm of expressions (\ref{PDF-fin})
for the distribution functions, one can obtain the free energy of a Gaussian chain in the external
parabolic potential. For the practical applications, it is convenient to combine both limiting cases
of small and large expansion factor into one extrapolating expression (see, e.g., \cite{khokhlov1994})
\begin{equation}
\label{Fren}
  \frac{F(R_g^2,\lambda)}{k_BT}  = -\ln P(R_g^2)
  \approx  \left[ \frac{3d}{4\alpha^2} +  \frac{\pi^2 d \alpha^2}{12} \right] + \frac{d\alpha^2\lambda^2}{6} ,
\end{equation}
where the expression in the square brackets refers to the free energy of a field-free chain. As it follows from Eq.~(\ref{Fren}), the contribution 
to the free energy from the interaction with the external field is additive and scales as
$\sim (\Omega R_g)^2$ for the quadratic potential $V(r) \sim \Omega^2 r^2$. 
This contribution can be obtained from a qualitative
analysis, but here it is found rigorously with the according coefficient. 
Note that this result may be also used as a
starting point for various empirical estimates of the free energy contribution associated 
with an external field.

\section{Comparison of the theory and Monte Carlo simulation results}
\label{sect:numres}

We applied  Monte Carlo simulations to obtain numerically the PDF $P(R_g^2)$ of the gyration radius
for different dimensions $d$.
We compared the numerical results with the theoretical predictions (\ref{PDF-case-1}) and
(\ref{PDF-case-2}) for different values of $\lambda$, which characterizes
the strength of the external field. In
simulations we used the Gaussian polymer with $N=100$ monomers.
All the parameters of the system, that is, the Kuhn's
length $b$ and thermal energy $k_{\mathrm{B}}T$ were set to be unit.
The parameter $\mathbf{r}_0$ in the harmonic potential~(\ref{V-field-potential}) was taken equal to zero.

The Monte Carlo method for calculating the PDF of the gyration radius has been
implemented as follows. At each
simulation we fixed the starting point at the origin and generated
random walks on a lattice (see, e.g., \cite{KW08}).
As the  result, we obtained a ``polymer'' of  $N$ monomers with the properties of an ideal chain
as we did not take into account any interactions between the monomers.
For each simulation we calculated the gyration radius by the expression (see, e.g.,\cite{DE13})
\begin{equation*}\label{Rg-def}
  R^2_g = \frac{1}{2N^2} \sum_{i,j=1}^N \left(\mathbf{r}_i - \mathbf{r}_j \right)^2,
\end{equation*}
where $\mathbf{r}_i$ is the position of the $i$th monomer on a lattice. We found the gyration radius for
$10^6$ generated
polymer samples and collected the results for $R_g^2$ into ``bins''. In this way we obtained a
normalized density histogram that estimates the function $P(R_g^2)$.

To calculate the PDF of the gyration radius in the presence of an external field, we treated the generated polymer
conformations as states of a system. The probability for the system to be in a certain state $k$ is described by the
Boltzmann factor (see, e.g,~\cite{DE13}):
\begin{equation}\label{p-k-Boltzmann}
  p_k = \frac{e^{-\beta E_k}}{\sum_{k} e^{-\beta E_k}},
\end{equation}
where $\beta = k_{\mathrm{B}} T$ and $E_k$ is the energy of the state $k$. For the harmonic potential
(\ref{V-field-potential}) the energy of the state $k$ was calculated as
\begin{equation*}
  E_k
  = \frac{\Omega^2}{2} \int_0^N d \tau \, \mathbf{r}^2(\tau)
  = \frac{\Omega^2}{2} \sum_{i=1}^N (\mathbf{r}_i^{(k)})^2,
\end{equation*}
where the index $(k)$ denotes the monomers positions generated on the $k$th random walks simulation. For each
simulation or state $k$, we calculated its probability with the Eq. (\ref{p-k-Boltzmann}). Then we calculated the PDF
$P(R_g^2)$ by summing all probabilities (\ref{p-k-Boltzmann}) of the states with the values
of the gyration radius within a certain small interval $[R_g^2, R_g^2 + \Delta R_g^2]$.

\begin{figure}[ht!]
\centering
  a)\includegraphics[width = 0.47\textwidth]{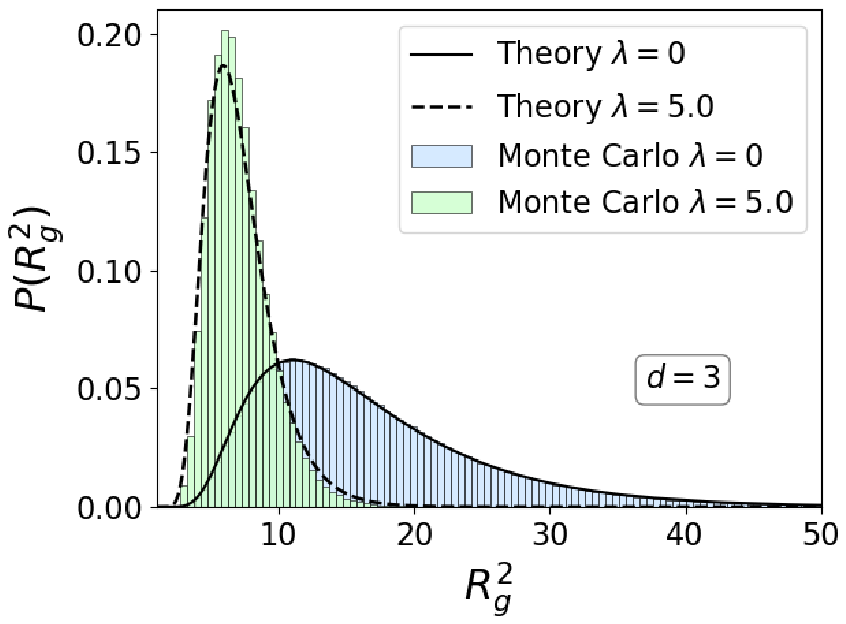}
  b)\includegraphics[width = 0.47\textwidth]{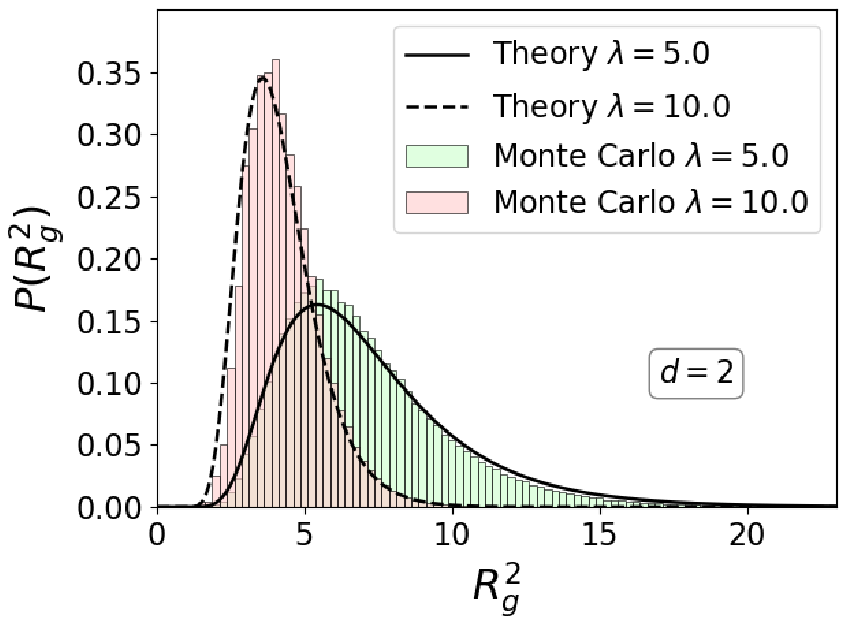}
  \caption{The PDF $P(R_g^2)$ of the gyration radius
  for the Gaussian polymer chains a) in three dimensions ($d=3$) and b) in two dimensions ($d=2$)
  for different values of the field parameter $\lambda$.
  The PDF is calculated by the
  Monte Carlo simulations and using Eq. (\ref{PDF-case-1}) in three dimensions for the field-free case
  ($\lambda = 0$) and Eq. (\ref{PDF-case-2})  in the presence of the field ($\lambda = 5.0$ and $10.0$).}
  \label{fig:PDF}
\end{figure}

In Fig.~\ref{fig:PDF}a, we compare the numerical and theoretical results for the PDF of the gyration radius $P(R_g^2)$ in three dimensions ($d=3$). For
the field-free case  ($\lambda = 0$), we compare our Monte Carlo numerical results with the
theoretical solution (\ref{PDF-case-1}) obtained in \cite{FN70},
where only the first four terms in the series are taken into account.
As can be seen from Fig.~\ref{fig:PDF}a,
our simulations results are in excellent agreement with the formula
(\ref{PDF-case-1}). This is also confirmed by the analysis of the moment
$\langle R_g^2 \rangle$ of the distribution function $P(R_g^2)$, presented in the Table \ref{tab:moments}.
Both theoretical and simulation results are very close to the exact value of
$\langle R_g^2 \rangle = N b^2/6 = 16.67$ (see
the second column in Table~\ref{tab:moments}).

For the strong external field ($\lambda = 5.0$ and $10.0$), our theoretical result (\ref{PDF-case-2})
for a small expansion factor, $\alpha \ll 1$, reproduces the simulation results quite accurately
(see~Fig.~\ref{fig:PDF}).
The investigation of $P(R_g^2)$ in the external field was carried out in $d=2,3,4$ dimensions.
The results in four dimensions do not differ much from $d=2$ and $3$ dimensions,
thus we do not present them here.
Since the saddle-point method gives an approximate solution,
it is convenient to keep the PDF normalized to unity by introducing the normalization constant $\mathcal{N}$ (see, e.g., \cite{Kle04}).
For $\lambda=5.0$, the calculated normalization
constant for the PDF (\ref{PDF-case-2}) reads, $\mathcal{N} = 5.3 \, (5.7)$ for $d=2 \, (3)$.
For $\lambda=10.0$ and $d=2$, we have $\mathcal{N} = 4.8$.
As one can see from Fig.~\ref{fig:PDF}a, the harmonic external field makes a
significant compaction of the polymer chain.
As $\lambda$ increases, the peak of $P(R_g^2)$ shifts to lower values of
$R_g$; accordingly, the height of $P(R_g^2)$ increases and its width decreases (see Fig.~\ref{fig:PDF}b).

The statistic characteristics of the PDF for three dimensions
are detailed in Table~\ref{tab:moments}. The Monte
Carlo simulations and theoretical formulas (\ref{PDF-case-1}) and (\ref{PDF-case-2})
give close results in the range of $\lambda$ from 0.0
to 10.0. The mean squared gyration radius $\langle R_g^2 \rangle$,
or the position of $P(R_g^2)$ peak, decreases almost four times from $16.67$ at
$\lambda = 0$ down to about $4.2$ at $\lambda = 10.0$ (see the first row in the table).
The variance $\sigma_{R_g^2}$ also decreases from about $8.5$ down to approximately $1.0$
(see the second row of the table). Thus, the PDF of the gyration
radius shrinks with the increasing  strength of the field.

\begin{table}[ht!]
\centering
\caption{The first moment $\langle R_g^2 \rangle$ and variance
$\sigma_{R_g^2}$ of the PDF $P(R_g^2)$ of the gyration radius
calculated in $d=3$ dimensions
by Monte Carlo simulations and using Eqs. (\ref{PDF-case-1}) and (\ref{PDF-case-2})
for different values of the external field parameter $\lambda$.}
\label{tab:moments}
\begin{tabular}{c|cc|cc|cc|cc}
\toprule
                        & \multicolumn{2}{c|}{$\lambda = 0$ (no field)} & \multicolumn{2}{c|}{$\lambda = 5.0$} & \multicolumn{2}{c|}{$\lambda = 7.5$} & \multicolumn{2}{c}{$\lambda = 10.0$} \\
\midrule
                        & MC                   & Theory               & MC              & Theory           & MC              & Theory           & MC             & Theory          \\
$\langle R_g^2 \rangle$ & 16.67                & 16.61                  & 7.05            & 7.19               & 5.38            & 5.27               & 4.34           & 4.17              \\
$\sigma_{R_g^2}$        & 8.47                 & 8.32                   & 2.27            & 2.55               & 1.46            & 1.55               & 1.04          & 1.07      \\
\bottomrule
\end{tabular}
\end{table}

\section{Conclusion}
\label{sect:conclusion}

We have investigated the probability distribution function (PDF) of the gyration radius $P(R_g^2)$ of a $d$-dimensional
Gaussian polymer chain in an external harmonic  potential. We have exploited 
the flexibility of the path integral technique to derive
an explicit expression for the characteristic function of the PDF.


We have obtained approximate expressions of $P(R_g^2)$ for small and large values of the expansion factor 
and Flory-type approximation for the free energy as a function of the gyration radius 
and field strength. The
latter quantities are important in thermodynamic perturbation theories, as the Gaussian chain 
is a very popular
reference system. We wish to stress that it is not straightforward (and most likely not possible) to find 
these expressions using the standard Fixman approach.

We have calculated the PDF $P(R_g^2)$ numerically using Monte Carlo simulations and found
that our theoretical expression for $P(R_g^2)$ reproduces the simulation results accurately when the external field is
strong. The first moment and variance of $P(R_g^2)$ calculated by simulations and theoretical formula agree well. As
the external field increases, the peak of $P(R_g^2)$ becomes sharp and shifts to lower values of $R_g$, this agrees
with the fact that the harmonic field causes the polymer to condense. The statistical description of polymer chains
with a more complicated potential rather than harmonic remains a subject for future researches.


\section*{References}
\bibliography{PGB2020JPhysA}

\end{document}